\documentclass[preprint,epsfig,aps,showpacs]{revtex4}
\usepackage{epsfig}
\begin{document}

\title{High strangeness dibaryons
 in the extended quark delocalization, color screening model}

\author{Hourong Pang}
\affiliation{Department of Physics,
Nanjing University, Nanjing, 210093, P. R. China;\\
Institute of Theoretical Physics, Chinese Academy of Sciences,
Beijing, 100080, China}
\author{Jialun Ping}
\affiliation{Department of Physics, Nanjing Normal University, 
Nanjing,
210097, P.R. China;\\
Center for Theoretical Physics, Nanjing University, Nanjing,
210093, P.R. China}
\author{Fan Wang}
\affiliation{Center for Theoretical Physics and Department of 
Physics,
Nanjing University, Nanjing, 210093, P. R. China}
\author{T. Goldman}
\affiliation{Theoretical Division, Los Alamos National 
Laboratory,
Los Alamos, NM 87545, USA}
\author{Enguang Zhao}
\affiliation{Institute of Theoretical Physics, Chinese Academy 
of Science,
Beijing, 100080, China}

\begin{abstract}
Dibaryon candidates with strangeness $S=-2,-3,-4,-5,-6$ are 
studied in terms of the extended quark delocalization and 
color screening
model. The results show that there are only 
a few promising low lying
dibaryon states: The $H$ and 
di-$\Omega$ may be marginally strong
interaction stable but 
model uncertainties are too large to allow
any definitive 
statement. The $SIJ=-3,1/2,2$ N$\Omega$ state is $62
MeV$ 
lower than the N$\Omega$ threshold and $24 MeV$ lower than
the $\Lambda\Xi\pi$ threshold. It might appear as a narrow 
dibaryon resonance and be detectable in the RHIC detector 
through the reconstruction of the vertex mass of the 
$\Lambda\Xi$ two body decay. The effects of explicit $K$ 
and $\eta$ meson exchange have been studied and found to be 
negligible in this model.
 The mechanisms of effective 
intermediate range attraction, $\sigma$ meson exchange and 
kinetic energy reduction due to quark delocalization are 
discussed.
\end{abstract}


\maketitle

\section{Introduction}

One of the most remarkable achievements of theoretical physics
in the past thirty years is the establishment and development
of the fundamental theory of the strong interaction -- quantum
chromodynamics (QCD). Perturbative QCD has been verified by
high energy experiments. However, the low energy physics of
QCD, such as hadron structure, hadron interactions and the
structure of exotic quark-gluon systems, are much harder to
calculate directly from QCD. One needs effective theories and
phenomenological models in these cases.

Borrowing the idea of quasiparticles from condensed matter and
nuclear physics, one can approximately transform the complicated
interactions between current quarks into dynamic properties of
quasiparticles and what is left to be studied are the residual
interactions between quasiparticles. One of the quasiparticles in
QCD is the constituent quark. How to dress the current quark 
to be a constituent quark still poses a theoretical challenge;
various effective theories\cite{cahill,tandy,roberts} have been
developed to derive constituent quarks from QCD. The common 
point of view is that the dynamical generation of the constituent 
quark mass is closely related to spontaneous chiral symmetry 
breaking initiated by the formation of a $q\bar{q}$ condensate 
in the QCD vacuum.

The constituent quark model has been quite successful in
understanding hadron spectroscopy and hadron interactions even
though we have not yet derived the constituent quark model
directly from QCD. There are also various versions of the
constituent quark model based on different effective degrees
of freedom. De Rujula, Georgi and Glashow\cite{prd12} first 
put
forward a quark-gluon coupling model based on constituent
quark and gluon effective degrees of freedom. Isgur obtained 
a
good description of hadron spectroscopy based on this
model\cite{isgur}. However, extension of the model to baryon
interactions does not reproduce the nucleon-nucleon
intermediate and long range interaction.

One modification studied is the addition of Goldstone boson
exchange on the quark level\cite{faesslerplb124,obukhoplb238,
fernjpg19,fujiprl}, in which the short-range part of the
interaction is described by the quark-gluon degrees of freedom
and the medium and long-range parts are attributed to
meson-exchange. This quark gluon-meson exchange hybrid model
achieves a quantitative fit of the nucleon-nucleon (NN) and
nucleon-hyperon (NY) scattering data.

A different modification of the De Rujula-Georgi-Glashow-Isgur
(RGGI) model, the quark delocalization and color screening model
(QDCSM)\cite{prl69}, has also been suggested. It maintains the
Isgur Hamiltonian for single hadrons but modifies it for
baryon-baryon (BB) interactions with two new ingredients.

First, the two center single quark orbital wave function (WF)
used in the quark cluster model is replaced by a delocalized
quark orbital WF. The introduction of quark delocalization can
be viewed either as taking into account the contribution of
excited configurations or the distortion of each individual
baryon due to their mutual interaction. This straightforward
method enlarges the variational Hilbert space. Its advantage
is that it permits the six-quark system to choose a more
favorable configuration through its own dynamics, while
maintaining a tolerable level of computational complexity.

Second, a different parametrization of the confinement
interaction is assumed, in which the usual quadratic confinement
and a color-screened quadratic confinement are used to
parameterize the two body matrix elements of different quark
orbits. The introduction of this parametrization is aimed at
taking into account some of the nonlinear, nonperturbative
properties of QCD which can not be described by two-body
quark-quark interactions in multiquark systems, such as the
formation of color flux tubes connecting many quarks, the three 
gluon interaction and the three body instanton interaction.

The QDCSM not only justifies the view of a nucleus as,
approximately, a collection of nucleons rather than a
single big bag with 3A quarks, but it also explains the long 
standing fact that the nuclear force and molecular forces 
are similar except for the obvious energy and length scale 
differences. It is also the model which requires the
fewest adjustable parameters to fit the existing BB interaction
data\cite{prl69,prc53,npa673,prc62,prc65,prc65p,lucpl03}.

There have been debates on which constituent quark model and
which effective degrees of freedom are best to use for hadron
structure and interaction studies\cite{prc57,kfliu,isgurprd6162,
georgiprd59,valcarceprc616364}. In our phenomenological study
of BB interactions with three different constituent quark models,
we found that even though the QDCSM and the other two models
appear to be quite different, they give similar BB interactions
in 44 of the 64 lowest BB channels consisting of octet and
decuplet baryons\cite{prc65}. A preliminary analysis
of the origin of this surprising similarity has been
produced\cite{nucl-th/0212012}. This result also implies that
quark delocalization and color screening, working together, do
provide the intermediate range attraction described by meson
exchange in other models. On the other hand, the different models
do give characteristically different results in some channels.
For example, the QDCSM predicts a strong attraction in the I=0,
J=3 channel, producing a dibaryon resonance, $d^*$, in this
channel, while the quark gluon-meson exchange hybrid model
predicts a strong attraction in the I=0, J=0 $\Omega\Omega$
channel, implying the existence of a strong interaction stable
di-$\Omega$. One cannot expect scattering data to become
available in these channels to test these model predictions,
but the dibaryon states should be detectable to provide a check
on whether these model predictions are realistic.

The study of dibaryon states not only checks constituent 
quark
models but also searches for new hadronic matter. 
The
$H$-particle has been assumed to be a six quark system 
from
the very beginning\cite{jaffe} and has been both a 
theoretical
and experimental topic for a long time. The 
S=0, $J^p =0^-$ d' dibaryon was assumed to be an NN$\pi$ 
system and was a
hot topic in the 1990's\cite{bilger,prc62}. 
We showed that the
$S=0$, $I=0$, $J=3$ $d^*$\cite{prc39,
prl69,prc51,mpla13,npa688,prc65,prc65p} is a tightly bound 
six quark system
rather than a loosely bound nuclear-like 
system of two
$\Delta$'s. An $S=-3, I=1/2, J=2$ N$\Omega$ 
state was proposed as a high strangeness
dibaryon 
candidate\cite{goldman}. Kopeliovich predicted high
strangeness 
dibaryons, such as the
di-$\Omega$ with $S=-6$, using the 
flavor SU(3) Skyrmion
model\cite{kopelnpa639}, and Zhang 
suggested searching for the
di-$\Omega$ in ultrarelativistic 
heavy
ion collisions\cite{065204}. Lomon predicted a 
deuteron-like
dibaryon resonance using R-matrix
theory\cite{lomon} and measurements at Saclay seem to offer
experimental support\cite{Sacley} for its existence.

In this paper, we calculate promising dibaryons with strangeness
$S=-2,-3,-4,-5$ using the extended quark-delocalization
color-screening model to provide another reference spectrum for
strange dibaryons and a possible further check of constituent
quark models.
 The SIJ=003 $d^*$ and -600 di-$\Omega$ dibaryons,
which we considered previously, are included in the discussion
to demonstrate the differences between the predictions of the
extended QDCSM and other quark models.

The extended QDCSM is briefly introduced in Section II. In
Section III, we present our results. We discuss these results
further in Sections IV and V, and conclude in Section VI.

\section{Brief description of the extended QDCSM}

The QDCSM was put forward in the early 90's. Details can
be found in Refs.\cite{prl69,prc51,npa688}. Although the
intermediate range attraction of the $NN$ interaction is
reproduced by the combination of quark delocalization and
color screening, the effect of the long-range pion tail
is missing in the QDCSM. Recently, the extended QDCSM 
was developed\cite{prc65p}, which incorporates this long-range
tail by adding $\pi$-exchange with a short-range cutoff.
The extended QDCSM not only reproduces the properties of the
deuteron well, but also improves agreement with NN scattering
data as compared to previous work\cite{lucpl03}.

The Hamiltonian of the extended QDCSM, wave functions and the
necessary equations used in the current calculation are given
below. Here we do not take into account the effect of any tensor
forces. The details of the resonating-group method (RGM) have
been presented in Refs.\cite{npa688,Buchmann}.

The Hamiltonian for the 3-quark system is the same as the well
known quark potential model, the Isgur model. For the six-quark 
system,
we assume 
\begin{eqnarray}
H_6 & = & \sum_{i=1}^6 (m_i+\frac{p_i^2}{2m_i})-T_{CM}
+\sum_{i<j=1}^{6}
    \left[ V_{conf}(r_{ij}) + V_G(r_{ij}) +V_{\pi}(r_{ij})
    \right] ,                    \nonumber \\
V_G(r_{ij}) & = & \alpha_s \frac{\vec{\lambda}_i \cdot
\vec{\lambda}_j }{4}
 \left[ \frac{1}{r_{ij}}-\frac{\pi}{2} \delta (\vec{r_{ij}})
 \left( \frac{1}{m^2_i}+\frac{1}{m^2_j}+\frac{4\vec{\sigma}_i
 \cdot \vec{\sigma}_j}{3m_im_j} \right) \right], \nonumber  \\
V_{\pi}(r_{ij}) & = & \theta (r-r_0)
\frac{g_8^2}{4\pi}\frac{m_{\pi}^2} {12m_q^2} \frac{1}{r_{ij}}
e^{-m_{\pi} {r_{ij}}} \vec{\sigma}_i \cdot \vec{\sigma}_j\vec{\tau}_i
\cdot \vec{\tau}_j,
   \label{hamiltonian} \\
V_{conf}(r_{ij}) & = & -a_c \vec{\lambda}_i \cdot \vec{\lambda}_j
\left\{ \begin{array}{ll}
 r_{ij}^2 &
 \qquad \mbox{if }i,j\mbox{ occur in the same baryon orbit}, \\
 \frac{1 - e^{-\mu r_{ij}^2} }{\mu} & \qquad
 \mbox{if }i,j\mbox{ occur in different baryon orbits},
 \end{array} \right. \nonumber \\
\theta ({r_{ij}}-r_0) & = & \left\{
 \begin{array}{ll}  0 & \qquad r_{ij} < r_0, \\  1 & \qquad \mbox{otherwise},
 \end{array} \right. \nonumber
\end{eqnarray}
where $r_0$ is the short range cutoff for pion exchange between
quarks. All the symbols have their usual meanings, and the
confinement potential $V_{conf}(r_{ij})$ has been discussed in
Refs.\cite{prc65p,npa688}.

The pion potential, $V_{\pi}(r_{ij})$, affects only the $u$
and $d$ quarks. We take these to have a common mass, $m_q=
m_d = m_u$, ignoring isospin breaking effects as they are
small on the scale of interest here.

The quark wave function in a given nucleon (orbit) relative
to a reference center (defined by $\vec{S}$) is taken to
have a Gaussian form characterized by a size parameter, $b$,
\begin{equation}
\phi(\vec{r}-\vec{S})  =  \left( \frac{1}{\pi b^2} \right)^{3/4}
    e^{-\frac{1}{2b^2} (\vec{r} - \vec{S})^2} .  \label{qkwvfcn}
\end{equation}

The values of $m_q$, $m_s$, $b$, $\alpha_s$ and $a_c$ are
determined by reproducing the $\Delta-N$ mass difference,
the nucleon mass, a hyperon mass and by requiring a stability
condition. The quark-pion coupling constant ${g_{qq\pi}}$
is obtained from the nucleon-pion coupling constant by a
slight ($<10$\%) correction to the classic symmetry relation,
viz.,
\begin{equation}
\frac{g^2_{NN\pi}}{4\pi} = (M_N/m_q)^2
    \left(\frac{5}{3}\right)^2 \frac{g^2_{8}}{4\pi} e^{m_{\pi}^2 b^2/2}
\end{equation}
where $M_N$ is the nucleon mass and the last factor provides
the correction due to the extent of the quark wavefunction in
the nucleon. The color screening parameter, $\mu$, has been
determined by matching our calculation to the mass of the
deuteron. All of the parameters are listed in Table I.

\vspace{12pt} Table I: Model Parameters

\begin{tabular}{ccccccc} \hline\hline
$m_q, m_s(MeV)$&$b(fm)$&$a_c(MeV\cdot
fm^{-2})$&$\alpha_s$&$\frac{g^2_{8}}{4\pi}$&~~~$~~r_0(fm)$~~~~~~&$\mu(fm^{-2})$ \\ 
\hline
$313, 634$&~~~$0.6015$~~~&~~~$25.14$~~~& ~~~$1.5585
$~~~&$0.5926$&~~~$0.8$~~~&~~~$0.85$~~~
\\ \hline\hline
\end{tabular}

\vspace{12pt}

The model masses of all octet and decuplet baryons are listed
in Table II.

\vspace{12pt} Table II: Single Baryon Masses in Units of MeV

\begin{tabular}{ccccccccc} \hline\hline
&N&$\Sigma$&$\Lambda$&$\Xi$&$\Delta$&$\Sigma ^*$&$\Xi^*$&$\Omega$ \\ \hline
theor.&~~~ 939.0~~~&1210.6~~&~~1113.6~~&1350.0&~~1232.0~~
&~~1358.1~~&~~1497.5~~&~~1650.1~~ \\ 
expt.&~~~ 939~~~&1193~~&~~1116~~&~~1318~~&~~1232~~
&~~1385~~&~~1533~~&~~1672~~ \\ \hline\hline
\end{tabular}

\vspace{12pt} \noindent
We use the resonating group method to carry out a
dynamical calculation. Introducing Gaussian functions with
different reference centers $S_i$ i=1...n, which play the role of
generating coordinates in this formalism, to expand the relative
motion wave function of the two quark clusters, we have
\[
\chi{ (\vec R)} = (\frac{3}{2\pi b^2})^{3/4} \sum_i C_i
e^{-\frac{3}{4}(\vec R - \vec S_i )^2/b^2}.
\nonumber
\]

In principle, any set of base wave functions can be used to
expand the relative motion wave function. The choice of a
Gaussian with the same size parameter, $b$, as the single
quark wave function given in Eq.(\ref{qkwvfcn}), however,
allows us to rewrite the resonating group wave function as
a product of single quark wave functions; (see Eq.(\ref{multi})
below). This cluster wave function (physical basis) can be
expressed in terms of the symmetry basis, classified by the
symmetry properties, in a group chain which in turn allows
the use of group theory methods to simplify the calculation
of the matrix elements of the six quark Hamiltonian\cite{wpg}. 
In our calculations, we typically use 12 Gaussian functions 
to
expand the relative motion wave function over the range
0-8~fm. For a few channels, such as the deuteron and $H$ 
particle, 20 Gaussian
functions are used to extend the range 
to 12~fm.

After including the wave function for the center-of-mass motion,
the ansatz for the two-cluster wave function used in the RGM can
be written as
\begin{eqnarray}
\Psi_{6q} & = & {\cal A}  \sum_k \sum_{i=1}^{n} C_{k,i}
  \int d\Omega_{S_i}
  \prod_{\alpha=1}^{3} \psi_{R} (\vec{r}_{\alpha},\vec{S_i},\epsilon)
  \prod_{\beta=4}^{6} \psi_{L} (\vec{r}_{\beta},\vec{S_i} , \epsilon)  \nonumber \\
  & &  [\eta_{I_{1k}S_{1k}}(B_{1k})\eta_{I_{2k}S_{2k}}(B_{2k})]^{I,J=S}
   [\chi_c(B_1)\chi_c(B_2)]^{[\sigma]}
    \label{multi}  ,
\end{eqnarray}
where $k$ is the channel index. For example, for $SIJ=-2, 0, 0$,
we have $k=1, 2, 3$, corresponding to the channels
$\Lambda\Lambda$, N$\Xi$ and $\Sigma\Sigma$.

The delocalized orbital wavefunctions, $\psi_{R}(\vec{r},\vec{S_i}
,\epsilon)$ and $\psi_{L}(\vec{r},\vec{S_i} ,\epsilon)$, are given by
\begin{eqnarray}
\psi_{R}(\vec{r},\vec{S_i}, \epsilon) & = & \frac{1}{N(\epsilon)}
    \left( \phi(\vec{r}-\frac{\vec{S_i}}{2}) + \epsilon
    \phi(\vec{r}+\frac{\vec{S_i}}{2}) \right) , \nonumber \\
\psi_{L}(\vec{r},\vec{S_i}, \epsilon) & = & \frac{1}{N(\epsilon)}
    \left(\phi(\vec{r}+\frac{\vec{S_i}}{2}) + \epsilon
    \phi(\vec{r}-\frac{\vec{S_i}}{2})\right) , \label{1q} \\
N(\epsilon) & = & \sqrt{1+\epsilon^2+2\epsilon e^{-S_i^2/4b^2}},
\nonumber
\end{eqnarray}
where $\phi(\vec{r}-\frac{\vec{S_i}}{2})$
and $\phi(\vec{r}+\frac{\vec{S_i}}{2})$
are the single-particle Gaussian quark wave functions referred
to above in Eq.(\ref{qkwvfcn}), with different
reference centers
$\frac{{S_i}}{2}$ and $-\frac{{S_i}}{2}$, respectively. The
delocalization
parameter, $\epsilon$, is determined by the
dynamics of the quark system
rather than being treated as an adjustable parameter.

The initial RGM equation is
\begin{equation}
\int H(\vec R, \vec{R'}) \chi (\vec{R'}) d\vec{R'} = E \int
N(\vec R,\vec{R'}) \chi (\vec{R'}) d\vec{R'} .
\label{RGM}
\end{equation}
With the above ansatz, the RGM Eq.(\ref{RGM}) is converted
into an algebraic
eigenvalue equation,
\begin{equation}
\sum_{j,k'} C_{j,k'} H_{i,j}^{k,k'}
  = E \sum_{j} C_{j,k} N_{i,j}^{k},
   \label{GCM}
\end{equation}
where $N_{i,j}^{k}, H_{i,j}^{k,k'}$ are the wave function
overlaps and Hamiltonian matrix
elements, respectively,
obtained for the wave functions of Eq.(\ref{multi}).

\section{Results and Discussion}

Previously, we chose the di-$\Omega$ as an example to study
whether or not our model results were sensitive to the
meson-exchange cut-off parameter, $r_0$, and the result
demonstrates that they are not\cite{commun38}. Hence, we consider
it sufficient to calculate six-quark systems of different
strangeness, $S=-2,-3,-4,-5$, with a representative cutoff value
of $r_0=0.8 fm$. Table III displays the masses (in MeV) calculated
for the strange dibaryon states of interest here. The lowest
(without taking tensor coupling into account) channel for each SIJ
is identified by bold lettering; $sc$ and $cc$ denote single
channel (the lowest one) and coupled channels, respectively.
'*' denotes those states without coupled channels. Additionally, 
it should be noted that in our calculation we assume the wavefunction 
to be
zero at the boundary point, 
which is the usual boundary 
condition for bound states. If the
state is unbound, we will not 
obtain a stable minimum eigenenergy in the course of extending 
the boundary point.
The unbound states are denoted by "-" in 
Table III.

\begin{center}
Table III : Masses of Six-Quark Systems with Strangeness
\begin{tabular}{cccc}  \hline\hline
S,I,J &coupling channels& Mass$_{sc}$&Mass$_{cc}$ \\
\hline $-2,0,0$ & ${\bf
\Lambda\Lambda}$-$N\Xi-\Sigma\Sigma$ &-- &$2225.5 $
\\ 
$-3,1/2,2$ & ${\bf N\Omega}-\Sigma\Xi^*-\Xi\Sigma^*-\Xi^*
\Lambda-\Xi^* \Sigma^*$ & $2566.4 $&$2549.1$ \\
  $-3,1/2,1$& ${\bf \Lambda \Xi}
-N\Omega-\Lambda\Xi^*-\Xi\Sigma^*-\Sigma^* \Xi^* - \Sigma\Xi^* -
\Sigma\Xi$& --  & -- \\  $-4,1,0$ & ${\bf
\Xi\Xi}-\Sigma^* \Omega-\Xi^* \Xi^*$ & -- & --
\\
$-4,0,1$ &  ${\bf \Xi\Xi} - \Lambda \Omega-\Xi\Xi^*-\Xi^*\Xi^* $ &
 --&  --    \\ 
$-5,1/2,0$ & ${\bf \Xi^* \Omega}$ & $3145.0 $&* \\
 $-5,1/2,1$ & ${\bf \Xi\Omega}-\Xi^* \Omega$ & -- & --
 \\ 
$-6,0,0$&${\bf \Omega\Omega}$&$ 3298.2$ &* \\ \hline\hline
\end{tabular}
\end{center}

In 1977, Jaffe\cite{jaffe} studied the color-magnetic 
interaction
of the one-gluon exchange potential in the
multiquark system and found that the most attractive
channel is the flavor singlet with quark content
$u^2d^2s^2$. Moreover, the same symmetry analysis
of the chiral boson exchange potential also leads to
the very same conclusion\cite{nucl-th/0212012}.

However dibaryon physics can be very delicate. The
deuteron channel is not a channel with strong attraction
in any baryon interaction model. If the deuteron had not
been found experimentally, it seems highly unlikely that
any model would have been able to predict it to be  a
stable dibaryon. The $H$-particle ($SIJ=-200$) is a six 
quark
state consisting mainly of octet-baryons, similar
to the deuteron, and we find only a weak attraction 
there in our model also. Hence, a qualitative analysis
is insufficient to judge whether or not the $H$-particle
is strong interaction stable. Systematically, we find
that a strong attraction develops only in decuplet-decuplet
channels and a mild attraction in octet-decuplet channels.

Moreover, in the $H$-particle case, the channel coupling effect 
may
even be more important than the deuteron case. In fact, it  
is not bound without taking coupled channels into account. In 
our
calculation, three channels have been taken into account. 
These
are: $\Sigma\Sigma$, N$\Xi$ and $\Lambda\Lambda$. 
The relative motion wave functions of each channel
contribution 
to the $H$-particle are shown in Fig 1. We find that
the  
$\Lambda\Lambda$ channel provides the largest contribution (67\%),
followed by the N$\Xi$ channel (23\%); the $\Sigma\Sigma$
channel 
contributes only (10\%). However, according to the
analysis 
by Jaffe, the biggest contribution is the $N\Xi$ channel, and
the $\Lambda\Lambda$ channel provides the smallest contribution.

The lowest mass we find for the $u^2d^2s^2$ system is $2225.5
MeV$, which is $6 MeV$ lower than the experimental threshold
of $\Lambda\Lambda$ and $1.7 MeV$ lower than our model threshold.
These values are smaller than the rms uncertainty that may be
inferred from our fit to the baryon octet and decuplet in
Table II. Furthermore, these values are on the order of corrections
one would expect from isospin violations which we have not
included. Hence, we can draw no definite conclusion as to whether 
or not the $H$-particle
is strong interaction stable in our model 
and we consider this
to be consistent with recent experimental 
findings\cite{tak}.

Besides the binding energy of the $H$, an interesting question
regarding the $H$ is its compactness, i.e., whether the $H$ is
a compact 6-quark object or a loosely bound $\Lambda\Lambda$
state. Fig.1 indicates that the maxima of the relative motion
wave function of the dominant $\Lambda\Lambda$ channel occurs 
around 1.8 fm, and the delocalization
parameter, $\epsilon$, of 
the dominant channel at the maximum
is $\sim$0.1; hence, the $H$ 
is a loosely bound system similar to the deuteron in our model.
Such a similarity may well be more physically sensible than a 
compact 6-quark structure. 

For systems with strangeness $S=-3$, we have
calculated the state N$\Omega$ (SIJ=-3,1/2,2), which was
shown to be mildly attractive, with energy below
$\Lambda\Xi\pi$ threshold\cite{goldman}. That conclusion
was challenged by Oka\cite{okaprd38} and supported by
Silvestre-Brac and Leandri\cite{slprd45}.

We have carried out a dynamical channel coupling calculation
to examine this state further. The $N\Omega, \Lambda\Xi^*,
\Xi\Sigma^*, \Sigma\Xi^*, \Xi^*\Sigma^*$ channels are all
included. We find the eigenenergy to be 2549.1 MeV, $24(54)
MeV$ lower than the $\Lambda\Xi\pi$ experimental(model)
threshold. Relative motion cluster wave functions for the
individual channels are shown in Fig. 2. The N$\Omega$
channel is by far the dominant one (77\%) and the maximum
for its relative motion wave function occurs at around 0.8 fm.
The value of $\epsilon$ at this separation is 0.46.
Mixing into the other channels is small. Hence, we find this  
to be a compact six quark state. 

Note that the D-wave $\Lambda\Xi$
and $\Sigma\Xi$ channels 
have not been included since no
tensor interaction has been 
included in this calculation.
This coupling should be weak 
because there is no $\pi$exchange in these channels and K 
and $\eta$ exchanges are not important in our model (see next 
section). Its effect on the eigenenergy of the
$N\Omega$ 
(SIJ=-3,1/2,2) state should be small and the D-wave decay 
widths to $\Lambda\Xi$ and $\Sigma\Xi$ final states should 
be small also. Therefore we expect the $SIJ=-3,1/2,2$
$N\Omega$ should be a narrow dibaryon resonance. The
tensor coupling calculation is currently in progress.

We have also calculated the state $SIJ=-3,1/2,1$. 
In the course 
of extending the boundary point, the lowest eigenenergy closely
approaches the model threshold but fails to come to a stable value 
within our limits of computation for the extension.
This result is unchanged by taking into account all possible 
coupling channels. Due to the fact that there are only weak 
attractions in our model for the octet-octet channels and the 
size of the model uncertainty, it is difficult to conclude whether 
or not there are strong interaction
stable states in these channels. 
We can only conclude that we have not found evidence for a strong 
interaction stable state with $SIJ=-3,1/2,1$.  

For systems with S=-4, we take the quantum numbers $SIJ=-4,1,0$ as
an example, because this case bears a number of similarities to
the deuteron. For example, in both cases, the lowest mass channel
is composed of two octet baryons from the same isodoublet. Also,
the matrix element $P^{sfc}_{36}$ characterizing the symmetry
property of the system is -1/81 for both cases. ($P^{sfc}_{36}$ is
the permutation operator of the quarks between two clusters acting
in spin, flavor and color space.) The result shows that the
system with S=-4, I=1, J=0 is unbound, even when the $\Xi^{*}\Xi^{*}$ 
and $\Sigma^{*}\Omega$ channel
couplings are taken into account. 
Because the calculated energy is again very close to
the $\Xi\Xi$ 
threshold this conclusion should be viewed as tentative.

For comparison, we have also calculated the $SIJ=-4,0,1$ state as
shown in Table III. The $\Xi\Xi$, $\Xi\Xi^*$, $\Lambda\Omega$ 
and $\Xi^*\Xi^*$ coupling channels are included. Our model result is 
very similar to the $SIJ=-4,1,0$, i.e., we do not find a bound state 
in this
channel. The $^3S_1-^3D_1$ tensor coupling should be small 
in the $\Xi\Xi$ channel so taking into account the tensor coupling
will not change the unbound character.

For the system with S=-5, we take the $SIJ=-5,1/2,0$ state as
an example. This state is interesting due to its $<P^{sfc}_{36}>$
value, $\sim -1/9$, which makes it a Pauli principle favored
state. If only two-baryon S-wave channels are taken into account,
there is only one channel for this state. Our calculation shows
that the contribution of the kinetic energy term, due to quark
exchange and delocalization effects, contributes strongly towards
the formation of a bound state. However, the one-gluon-exchange
interaction largely compensates for this attraction and produces
a mass of $3145.0 MeV$,
which is $59 MeV$ lower than the experimental 
value of
the $\Xi^*\Omega$ threshold but only about $2 MeV$ lower
than the model threshold. (The one-gluon-exchange effect here is
quite different from that in the $d^*$ case, where large
delocalization is favored for a wide range of cluster separations
as well as there being a strong effective attraction due to
the large reduction in the kinetic energy that accompanies
significant delocalization.) We conclude that this state is
not a good candidate for a dibaryon resonance search due to
its small binding and its $\Xi^{*}\Omega$ content.

Inclusion of the tensor interaction will mix the spin-2 D-wave
$\Xi\Omega$ channel with the spin-0 S-wave $\Xi^*\Omega$ 
channel. The tensor coupling should be weak also and we expect 
that its effect on the $SIJ=-5,1/2,0$ state is small.

In the same strangeness sector, we also calculated the
$SIJ=-5,1/2,1$ state, since it includes the lowest channel,
$\Xi\Omega$. The calculated energy is very close to the 
$\Xi\Omega$ threshold but a little higher so again there is 
no bound state, with $SIJ=-5,1/2,1$, in our model.

To sum up, there are only a few high strangeness states worthy
of experimental searches in our model. These are the $H$
particle, the N$\Omega$ and the di-$\Omega$. The di-$\Omega$ was
previously reported in Ref.(\cite{commun38}) and the result is
included in Table III.

The $H$-particle and the di-$\Omega$ may be strong interaction
stable. However in our model, the binding energies of both are
small relative to the model uncertainty. The di-$\Omega$ mass is
about $47 MeV$ lower than the experimental $\Omega\Omega$
threshold. However our model mass for the $\Omega$ is $1650 MeV$.
If this model mass of $\Omega$ were used to calculate the
threshold, then the di-$\Omega$ mass is no more than $2 MeV$ below
that threshold. Since our model mass for the single $\Omega$
baryon deviates from the experimental value about $22 MeV$, a
reasonable estimate of the model uncertainty for the dibaryon
would be at least that large. Therefore the di-$\Omega$ should not
be claimed as a strong interaction stable dibaryon within the
model. Similarly, we cannot claim that the $H$-particle is strong
interaction stable either. The $SIJ=-3,1/2,2$ N$\Omega$ case is
certainly not strong interaction stable. However, the state is
also certainly lower in mass than the N$\Omega$ threshold, and
quite possibly lower than the $\Lambda\Xi\pi$ threshold, as well. 
The tensor coupling to the $\Lambda\Xi$ and $\Sigma\Xi$ 
channels should be weak and the decay width should be small.
This strongly suggests that it is a promising candidate for a 
narrow dibaryon resonance. This prediction can be tested by
relativistic heavy ion reactions using RHIC detectors
through the reconstruction of the vertex mass of the two
body decay products, $\Lambda$ and $\Xi$.

\section{The {\bf $K,\eta$} meson exchange effect in the extended
QDCSM}

The effects of K and $\eta$ meson exchange have been studied in the
di-$\Omega$ channel and found to be negligible in our model\cite
{commun38}. In this section, we carry out a further systematic 
study of the effect of ($K,\eta$) meson exchange on the masses 
of strange dibaryon candidates. A flavor-symmetric octet 
meson-quark coupling is assumed for all of the octet mesons ($\pi, K, 
\eta$). The quark-meson exchange potential has the usual pseudoscalar 
meson exchange form except for a short range cut-off,
\[V_{K}(r_{ij})  = \sum_{a=4}^7 \theta (r-r_0)
\frac{g_8^2}{4\pi}\frac{m_{K}^2} {12m_im_j} \frac{1}{r_{ij}}
e^{-m_{K} {r_{ij}}} \vec{\sigma}_i \cdot \vec{\sigma}_j\vec{\lambda}^{f,a}_i
\cdot \vec{\lambda}^{f,a}_j, \]

\[V_{\eta}(r_{ij})  = \theta (r-r_0)
\frac{g_8^2}{4\pi}\frac{m_{\eta}^2} {12m_im_j} \frac{1}{r_{ij}}
e^{-m_{\eta} {r_{ij}}} \vec{\sigma}_i \cdot \vec{\sigma}_j\vec{\lambda}^{f,8}_i
\cdot \vec{\lambda}^{f,8}_j. \]
A unified cutoff of $r_0=0.8 fm$ is used. This is aimed at 
avoiding double counting since, in our model approach\cite{prc65},  
short and intermediate range interactions have been accounted 
for by the combination of quark delocalization and color screening. 
The model parameters ($b,a_c,\alpha_s$) are refitted as before, 
and are listed in Table IV. 
The recalculated masses of octet and 
decuplet baryons are listed
in Table V.

\vspace{12pt} Table IV: Model Parameters

\begin{tabular}{ccccccc} \hline\hline
$m_q, m_s(MeV)$&$b(fm)$&$a_c(MeV\cdot
fm^{-2})$&$\alpha_s$&$\frac{g_8^2}{4\pi}$&~~~$~~r_0(fm)$~~~~~~&$\mu(fm^{-2})$ \\ 
\hline
$313, 634$&~~~$0.6022$~~~&~~~$25.03$~~~& ~~~$1.5547
$~~~&$0.5926$&~~~$0.8$~~~&~~~$0.90$~~~
\\ \hline\hline
\end{tabular}

\vspace{12pt} Table V: Single Baryon Masses in Units of MeV

\begin{tabular}{ccccccccc} \hline\hline
&N&$\Sigma$&$\Lambda$&$\Xi$&$\Delta$&$\Sigma ^*$&$\Xi^*$&$\Omega$ \\ \hline
theor.&~~~ 939.0~~~&1217.47~~&~~1116.90~~&1357.56&~~1232.0~~
&~~1359.61~~&~~1499.70~~&~~1652.27~~ \\ 
expt.&~~~ 939~~~&1193~~&~~1116~~&~~1318~~&~~1232~~
&~~1385~~&~~1533~~&~~1672~~ \\ \hline\hline
\end{tabular}
 
\vspace{12pt}
Comparing Tables I,II and IV,V, it is apparent that the 
addition of K and $\eta$ exchanges has modified the model 
parameters and single baryon masses only slightly.
We also find that the properties of the deuteron are 
reproduced as well as before by a minor readjustment of 
the color-screening parameter $\mu$ \cite{commun38}.

With the full octet meson exchange, we recalculated every 
single channel
case with different quantum numbers. For the 
$H$-particle, we calculated
both the single channel and coupled 
channels, because it is very
sensitive to channel coupling.
The results are presented in Tables VI and VII. For comparison,
the results with $\pi$ exchange only are also listed. 
\begin{center}
Table VI : Masses of Six-Quark Systems with $\pi$ exchange only
and 
full octet pseudoscalar meson exchange. The masses are given in MeV.
\begin{tabular}{cccc}  \hline\hline
S,I,J &single channel&~~~ only $\pi$ exchange &~~~~~~$\pi,K,\eta$ exchange \\
\hline 
$-3,1/2,2$ & $N\Omega$ & $2566.37 $&$2556.95 $ \\
  $-3,1/2,1$& $\Lambda \Xi$& --  & -- \\  
$-4,1,0$ & $\Xi\Xi$ & -- & -- \\
$-4,0,1$ &  $\Xi\Xi$ & --&  --    \\ 
$-5,1/2,1$ & $\Xi\Omega $& -- & -- \\
 $-5,1/2,0$ & $\Xi^*\Omega$ & $3145.01$ & $3146.15$ \\ 
$-6,0,0$&$\Omega\Omega$&$ 3298.20$ &$3300.00$ \\ \hline\hline
\end{tabular}
\end{center}

\begin{center}
Table VII : Mass of $H$-particle ($SIJ=-2,0,0$) with $\pi$ exchange only
and full octet pseudoscalar meson exchange. The masses are given in MeV.
\begin{tabular}{cccc}  \hline\hline
&& only $\pi$ exchange &~~~~~~$\pi,K,\eta$ exchange \\
\hline 
&$\Sigma\Sigma$& $2280.40$ &$2282.67$ \\
single channels &$N\Xi$& $2263.95$& $2268.12$ \\
&$\Lambda\Lambda$& -- &-- \\ \hline
coupled channels & $\Sigma\Sigma-N\Xi-\Lambda\Lambda$ &$2225.48$ & $2230.28$ \\
\hline\hline
\end{tabular}
\end{center}

Tables VI and VII show that those systems which are unbound in the 
extended QDCSM with $\pi$ exchange only, such as $SIJ=-3,1/2,1; 
-4,1,0; -4,0,1$ and $-5,1/2,1$, remain unbound after adding 
$K$ and $\eta$ exchanges, while those systems, which are bound 
in Table III, remain bound. Moreover, their masses
are almost 
unaffected by the addition of $K$ and $\eta$ exchange.
The largest 
difference is not more than $10 MeV$.
Especially for the $H$-particle, 
the $\Lambda\Lambda$ channel would be unbound without
taking into 
account the $N\Xi$ and $\Sigma\Sigma$ channel
coupling; this character 
is not affected either by adding $K$ and $\eta$ exchange.
Even the $\Lambda\Lambda$, $N\Xi$ and $\Sigma\Sigma$ channel mixing
fractions, $56$\%, $23$\% and $21$\%, respectively, are very similar
to the case with $\pi$ exchange only.
 All of these results confirm 
our expectation that heavier meson ($K$, $\eta$)
exchange has already 
been mostly accounted for by the quark delocalization and color screening 
effects in our model approach. Hence, explicit inclusion of $K$ and 
$\eta$ exchanges beyond the cutoff scale is not important in our approach. 

\section{Further Discussion about the Mechanism of Intermediate Range
Attraction}

We have reported that the QDCSM gives very similar effective BB
interactions to other models\cite{prc65}, in general. However
a careful comparison found that our model results for the
binding energy of high strangeness dibaryons are systematically
smaller than those of the chiral quark model\cite{zhangzy}. 
The difference is mainly due to the different mechanism for the 
effective BB intermediate range attraction.

In the chiral quark model\cite{fernjpg19,fujiprl,zhangzy} the
intermediate range attraction is attributed to $\sigma$ meson
exchange. Because of its scalar-isoscalar character, $\sigma$
meson exchange provides a universal attraction independent 
of the flavor. 

In the Bonn meson exchange model\cite{mah1} the $\sigma$ meson is 
an effective
description of correlated two $\pi$ exchange. This 
point has been confirmed by a chiral perturbation calculation 
of the NN interaction\cite{mah2}. Such an effective $\sigma$-N 
coupling can not be extended from the NN channel to other
channels with strangeness by a universal $\sigma$-baryon
coupling, because in the NN channel, there are NN,
N$\Delta$ 
and $\Delta\Delta$ intermediate states for two $\pi$ exchange, 
while for the
$\Lambda\Lambda$ channel, there is only the 
$\Sigma\Sigma$
intermediate state, and for the $\Omega\Omega$
channel, no such intermediate state is possible. Therefore it 
is
not justified to fix the parameters of $\sigma$ exchange 
in
the NN channel and then directly extend that exchange to 
channels with strangeness.

There are arguments based on spontaneous chiral symmetry breaking
for introduction of $\sigma$ meson quark coupling. In the
SU(2) case, the non-linear realization of chiral symmetry can
be linearized and in turn the $\sigma$ and $\pi$ meson coupling
constants can be unified\cite{obukhoplb238,fernjpg19}. However one
should note this universal coupling is restricted to the $u$ and 
$d$ quarks and the effective $\sigma$ is still due to even number 
multiple $\pi$ exchange. In the SU(3) case, one can introduce 
SU(3) chiral symmetry by neglecting
the difference between the 
$s$ and light quark masses, followed by spontaneous chiral symmetry 
breaking.
However, the nonlinear realization of SU(3) can not be 
linearized
in the same way as in the SU(2) case to obtain a universal 
$u$, $d$ and $s$ quark $\sigma$ coupling.  The internal quark 
structure of the $\sigma$ meson is a controversial issue, but an equal 
mix of $u\bar{u}+d\bar{d}+s\bar{s}$ is quite unexpected and so also 
is a universal $u$, $d$ and $s$ quark $\sigma$ coupling.
Hence, a 
strong attraction in high strangeness channels arising
from such a 
universal $\sigma$ quark coupling is quite questionable also.

In the QDCSM, quark delocalization and color screening work 
together to provide appropriate short-range
repulsion and 
intermediate-range attraction for different channels.
We illustrate this mechanism by showing contributions 
of the kinetic energy, confinement, color Coulomb, and color 
magnetic terms to the effective BB potential $V_{BB}(S)$, 
as well as the total sum, in curves a-e, respectively in 
Figs.3-6 for a few typical channels. The value of $\epsilon$ 
varies with the separation $S$ and is also listed in Figs.3-6. The
contribution of $\pi, K, \eta$ exchange with a cut-off $r_0=0.8 fm$ 
is small so we do not show it. These figures are ordered in terms 
of increasing values of $<P^{sfc}_{36}>$, from Pauli favored to
Pauli forbidden.

The effective baryon-baryon interactions shown in these figures 
are obtained in the Born-Oppenheimer approximation, 
\[ V_{BB}(S_i)= \frac{\langle \Psi_{BB}(S_i) | H| 
\Psi_{BB}(S_i) \rangle }{\langle \Psi_{BB}(S_i) | 
\Psi_{BB}(S_i) \rangle } -\frac{\langle \Psi_{BB}(S_i)|H|\Psi_{BB}(S_i)\rangle}
{\langle \Psi_{BB}(S_i)|\Psi_{BB}(S_i)\rangle}|_{S_i\rightarrow\infty,
\epsilon=0}. \]
Where the $\Psi_{BB}(S_i)$ is the antisymmetric six quark cluster state 
at a specified separation, $S_i$, as given in Eq. (4) without summation 
over $k$ and $i$; $H$ is the six quark Hamiltonian (1).
The contribution of each term in Eq. (1) is defined similarly,
for 
example,
\[ V_{BB}^{conf}(S_i)= \frac{\langle \Psi_{BB}(S_i) | V_{conf}| 
\Psi_{BB}(S_i) \rangle }{\langle \Psi_{BB}(S_i) | 
\Psi_{BB}(S_i) \rangle } - 
\frac{\langle \Psi_{BB}(S_i)|V_{conf}|\Psi_{BB}
\rangle}{\langle \Psi_{BB}(S_i)|\Psi_{BB}(S_i)\rangle}
\mid_{S_i\rightarrow\infty,
\epsilon=0}. \]

Quark exchange alone induces a weak reduction in kinetic energy.
Quark delocalization enhances this kinetic energy reduction.
The kinetic energy
reduction is also dependent on the strangeness 
of the channels due
to the inverse quark mass dependence of the 
quark kinetic energy.
The higher the strangeness, the smaller the 
contribution of quark
kinetic energy and hence the smaller the 
reduction of the kinetic
energy due to delocalization (see 
Figs.3-6(a)). This makes the
intermediate range attraction weaker 
for the higher strangeness
channels and is the main reason that 
our model gives less binding
for the $S=-4,-5,-6$ states than does 
the chiral quark
model\cite{zhangzy}.

If the usual quadratic color confinement is used, it does not
contribute to the effective BB interaction. With the introduction 
of quark delocalization, the usual color confinement contributes 
an effective repulsion, as shown in Fig.7. The
color Coulomb has 
a similar behavior, as shown in Figs.3-6(c);
it produces almost 
no contribution to the effective BB interaction
without quark 
delocalization, and contributes an effective
repulsion with quark 
delocalization. These two terms working
together almost totally 
forbid quark delocalization. This
implies that the internal structure 
of the baryon is unaffected
by the mutual interaction. However, that 
is inconsistent with the observed
difference between the nucleon 
structure function in a
nucleus and in isolation as seen in deep 
inelastic lepton
scattering (EMC effect).

On the other hand, there is no compelling reason to assume that
the two body confinement potential is a good approximation for
a multi-quark system. At the very least, the three gluon and the
three body instanton interactions, which do not contribute to the
$q\bar{q}$ meson and $q^3$ baryon but do contribute to the
multi-quark system, have been omitted in every two body confinement
model. The color flux scenario revealed in the lattice QCD calculation 
of two and three quark systems raises questions regarding an additive 
two body confinement approximation.  To take these facts into account, 
the QDCSM reparametrizes
the confinement by introducing color 
screening\cite{prl69,prc53,
npa673,prc65p,prc51,npa688}. Figs.3-6(b) 
show that after the
introduction of color screening, the confinement 
term contributes
an additional attraction, which mainly reduces the 
repulsive core
of the effective BB interactions (See Fig.7). There 
is a systematic uncertainty in our model related to this term which 
is not yet quantified.

The color magnetic term generally contributes to a repulsive core
except for a few Pauli favored channels where it contributes an
additional intermediate range attraction so that these channels
develop a strong effective attraction. It is also dependent on the
strangeness of the channels: The higher the channel strangeness,
the weaker the color magnetic contribution, due to the inverse
strange quark mass in the color magnetic term. Figs.3-6(d) display
these results.

Altogether, these figures show that in the QDCSM the intermediate
range attraction is mainly due to reduction of the quark kinetic
energy, and the degree of reduction of the quark kinetic
energy 
is connected with the degree of quark delocalization;
the latter 
is determined by the competition between these
four terms and that 
competition is different in different
channels. We have mentioned 
before that in the $d^*$ channel this
competition produces a large 
quark delocalization for a wide range
of separations between two 
$\Delta$-like quark clusters so that
the kinetic energy acquires 
a correspondingly large reduction,
which in turn gives rise to a 
very strong attraction in this channel. (See
Fig.3). The $SIJ=-6,0,0$ 
has the same $<P^{sfc}_{36}>$ as that of the $SIJ=0,0,3$ channel and 
has a similar
opportunity to develop large quark delocalizations and 
strong
attraction. However, there the competition does not allow as 
large
a quark delocalization to develop and so the resulting attraction 
is
not as strong as in the $d^*$ channel. (See Fig.4.)

\section{Conclusion}

To sum up, we have carried out a dynamical calculation for 
the
most promising dibaryon candidates with high strangeness 
by using
the extended QDCSM.  Only a few high strangeness states 
recommend themselves for experimental
searches. These are the $H$ 
particle, the N$\Omega$ and the di-$\Omega$.  We recommend searching 
for the $(SIJ=-3,1/2,2)$  using $N\Omega\rightarrow\Lambda\Xi$ 
D-wave decay vertex mass reconstruction with the RHIC detectors 
as the case with the best likelihood. 

In the QDCSM, it is the quark dynamics
that controls the competition 
among the four terms in the Hamiltonian (1): the kinetic energy, the 
confinement, the color Coulomb, and the color magnetic, and which 
determines
the overall effective BB interactions. These are quite 
similar,
though not identical, to those of the quark-gluon-meson 
hybrid
model\cite{prc65} for the majority of BB channels. The exchange 
of explicit $K$ and $\eta$ mesons is not important in our model for 
mass estimates of strange dibaryons because of our short range cut-off. 
 
This model, which has the fewest parameters, describes
the properties 
of the deuteron and the existing NN,
N$\Lambda$ and N$\Sigma$ scattering 
data. Up to now it is the only model which gives an explanation of the 
long standing fact that the nuclear and molecular forces are similar in 
character despite the obvious length and energy scale differences and 
that nuclei are well described as a collection of A nucleons rather than 
3A quarks. In view of the
fact that the $H$-particle has not been 
observed experimentally, the
BB interaction in the $\Lambda\Lambda$ 
channel\cite{tak} predicted by this model may be a good approximation 
of the real world.  Further refinement is
possible by including more 
channel couplings and spin-orbit
and tensor interactions. 

Of course, the QDCSM is only a model of QCD.
The high strangeness 
dibaryon resonances may be a good venue
for determining whether the 
QDCSM mechanism for the intermediate
range attraction is more 
realistic than that of a universal $u$, $d$, $s$ quark-$\sigma$ 
meson coupling.

This work is supported by NSFC contract 90103018 and by the U.S.
Department of Energy under contract W-7405-ENG-36.
 F. Wang would 
like
to thank the ITP for their support through the visiting program.

\pagebreak
FIGURE
CAPTIONS

Fig.1 Relative motion wave functions of the coupled channels for
the SIJ=-2,0,0 state with eigenenergy 2225.5 MeV.

Fig.2 Relative motion wavefunctions of the coupled channels for
the SIJ=-3,1/2,2 state with eigenenergy 2549.1 MeV.

Fig.3 Contributions of kinetic energy, confinement, color Coulomb
and color magnetic terms (a-d respectively) to the effective
potential and the total (e) for the SIJ=0,0,3 $\Delta\Delta$
channel with $P^{sfc}_{36}=-1/9$. In each subfigure, the dotted
curve is for delocalization parameter $\epsilon$=0.0, dashed
curve for $\epsilon$=0.5, dashed-dotted curve for $\epsilon=1.0$;
the solid curve is for the self-consistent value of $\epsilon$
determined by the system dynamics.

Fig.4 The same as Fig.3 for the state $SIJ=-6,0,0$ $\Omega\Omega$
with $<P^{sfc}_{36}>=-1/9$.

Fig.5 The same as Fig.3 for the state $SIJ=0,0,1$ NN
with $<P^{sfc}_{36}>=-1/81$.

Fig.6 The same as Fig.3 for the state $SIJ=-2,0,0$ $\Lambda\Lambda$
with $<P^{sfc}_{36}>=0$.

Fig.7 The contribution of conventional quadratic color confinement
to the effective BB interaction. The solid curve is for
$\epsilon$=0.0, dotted curve for $\epsilon$=0.5 and the dashed
curve for $\epsilon$=1.0.

\pagebreak

\begin{figure*}[t]
\includegraphics[height=4.0in]{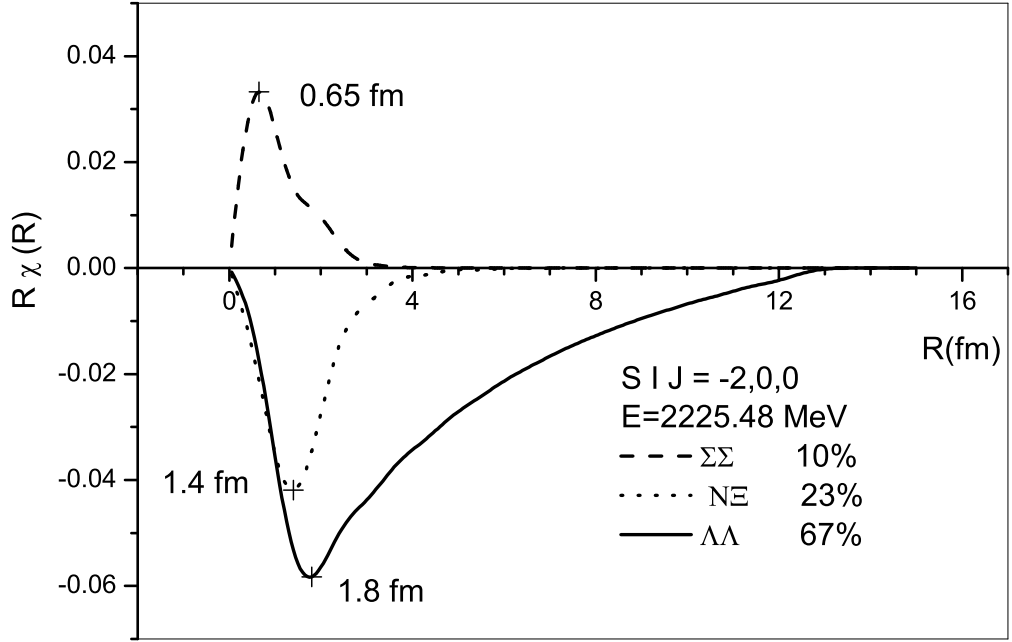}
\caption{Relative motion wave functions of the coupled channels for
the SIJ=-2,0,0 state with eigenenergy 2225.5 MeV.}
\label{FIG1}
\end{figure*}

\begin{figure*}[t]
\includegraphics[height=4.0in]{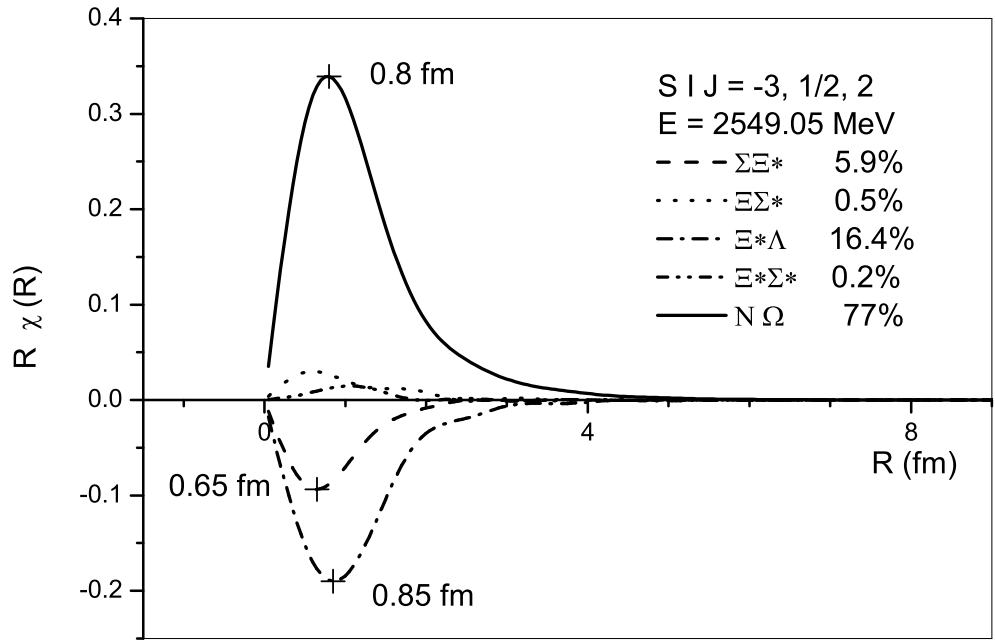}
\caption{Relative motion wavefunctions of the coupled channels for
the SIJ=-3,1/2,2 state with eigenenergy 2549.1 MeV.}
\label{FIG2}
\end{figure*}

\begin{figure*}[t]
\includegraphics[height=4.0in]{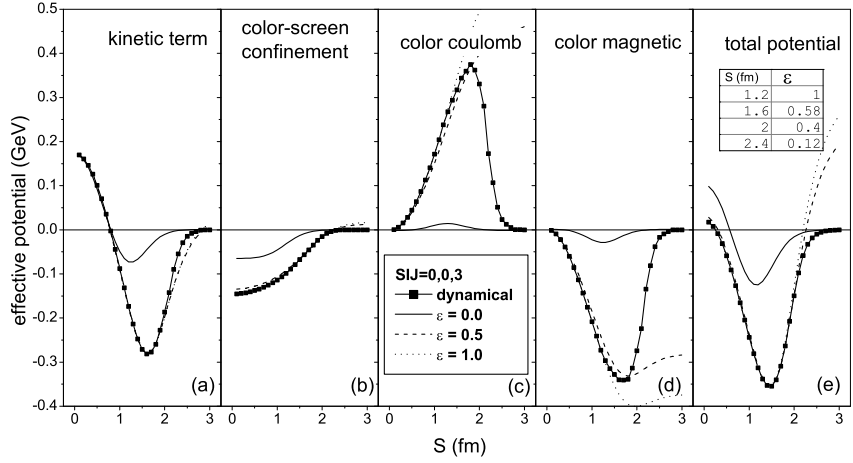}
\caption{Contributions of kinetic energy, confinement, color Coulomb
and color magnetic terms (a-d respectively) to the effective
potential and the total (e) for the SIJ=0,0,3 $\Delta\Delta$
channel with $P^{sfc}_{36}=-1/9$. In each subfigure, the dotted
curve is for delocalization parameter $\epsilon$=0.0, dashed
curve for $\epsilon$=0.5, dashed-dotted curve for $\epsilon=1.0$;
the solid curve is for the self-consistent value of $\epsilon$
determined by the system dynamics.}
\label{FIG3}
\end{figure*}

\begin{figure*}[t]
\includegraphics[height=4.0in]{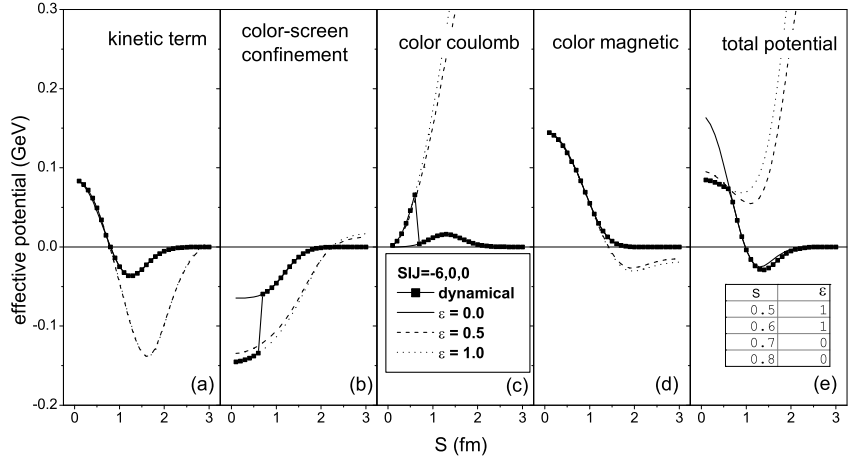}
\caption{The same as Fig.3 for the state $SIJ=-6,0,0$ $\Omega\Omega$
with $<P^{sfc}_{36}>=-1/9$.}
\label{FIG4}
\end{figure*}

\begin{figure*}[t]
\includegraphics[height=4.0in]{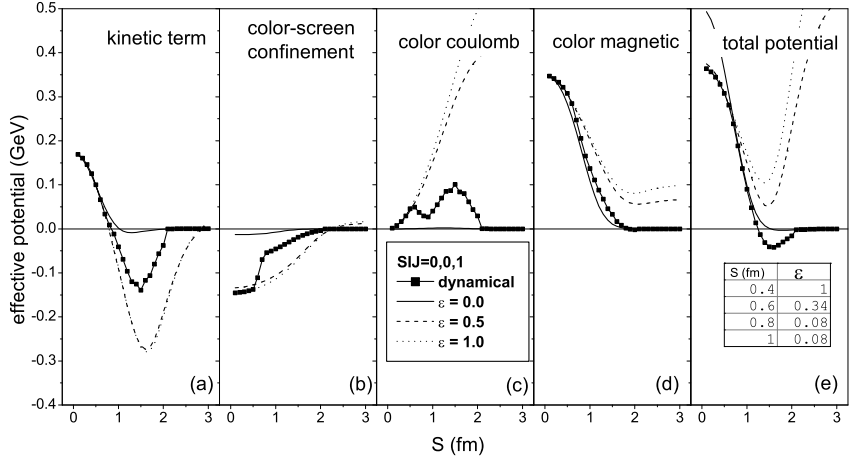}
\caption{The same as Fig.3 for the state $SIJ=0,0,1$ NN
with $<P^{sfc}_{36}>=-1/81$.}
\label{FIG5}
\end{figure*}

\begin{figure*}[t]
\includegraphics[height=4.0in]{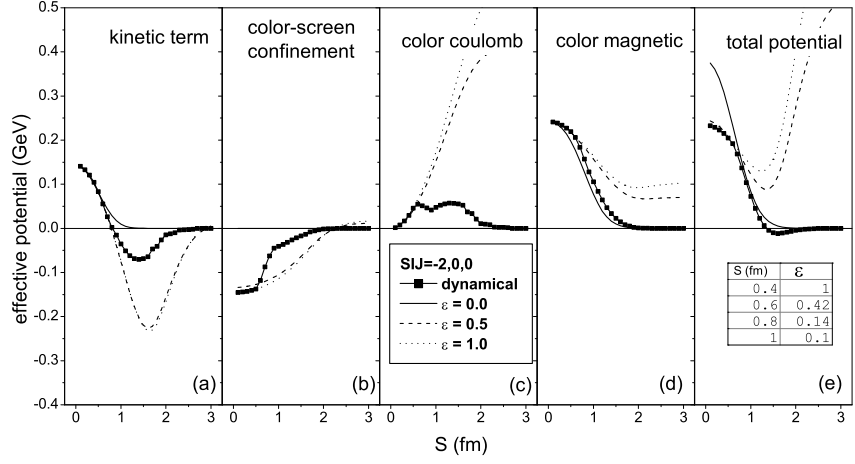}
\caption{The same as Fig.3 for the state $SIJ=-2,0,0$ $\Lambda\Lambda$
with $<P^{sfc}_{36}>=0$.}
\label{FIG6}
\end{figure*}

\begin{figure*}
\includegraphics[height=4.0in]{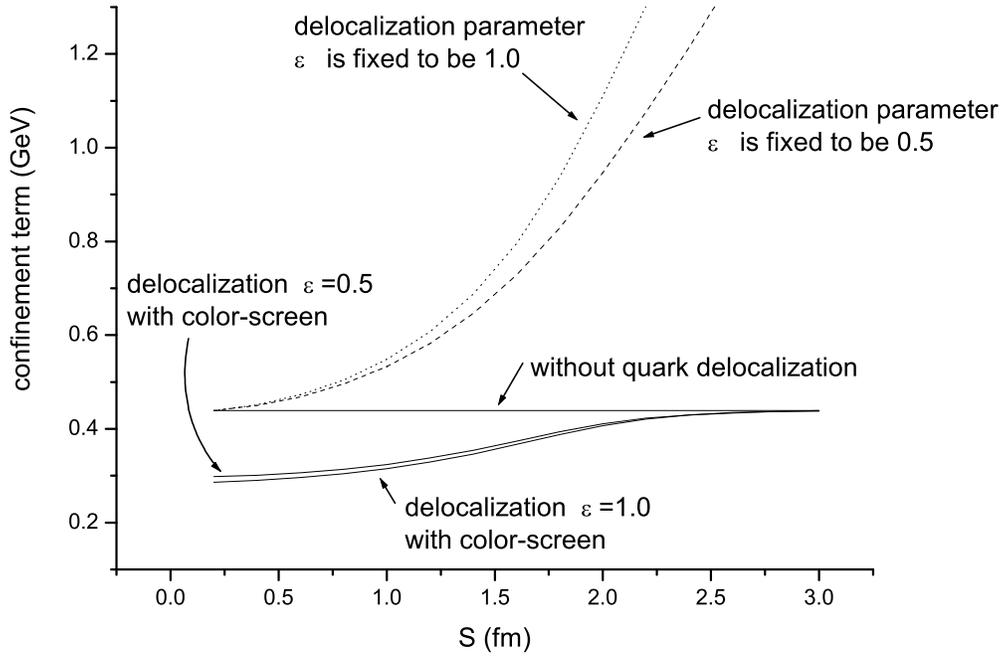}
\caption{The contribution of the usual quadratic color confinement
to the effective BB interaction with and without quark delocalization
and color-screening.} 
\label{FIG7}
\end{figure*}

\end{document}